\documentclass[aps,twocolumn,prc,superscriptaddress,showpacs,nofootinbib,floatfix,amssymb,amsfonts,amsmath]{revtex4-1}
\usepackage{graphicx}
\usepackage{epsf}
\usepackage{amsmath,amssymb}
\usepackage{bm}
\usepackage{color}
\usepackage{ulem}

\newcommand{\nuc}[2]{{}^{#2} \mathrm{#1}}
\begin{document}
\title{
Two $\nuc{Li}{9}$ clusters connected with two valence neutrons in $\nuc{C}{20}$
}% 

\author{Naoyuki Itagaki}
\author{Tokuro Fukui}
\affiliation{
Yukawa Institute for Theoretical Physics, Kyoto University,
Kitashirakawa Oiwake-Cho, Kyoto 606-8502, Japan
}

\author{Junki Tanaka}

\affiliation{
RIKEN Nishina Center for Aaccelerator-Based Science,
2-1 Hirosawa, Wako, Saitama 351-0198, Japan
}

\author{Yuma Kikuchi}
\affiliation{
Tokuyama College of Technology,
Gakuendai, Shunan, Yamaguchi 745-8585, Japan
}

\date{\today}

\begin{abstract}
Many preceding works have shown in $\nuc{Li}{11}$ the presence of the halo structure
comprised of the weakly bound two neutrons around $\nuc{Li}{9}$, 
and it is intriguing to see how this halo structure changes when another $\nuc{Li}{9}$ approaches.
In this study, we introduce a four-body model for $\nuc{C}{20}$
with two $\nuc{Li}{9}$ clusters and two valence neutrons.
The recent development of the antisymmetrized quasi cluster model (AQCM) makes it possible to
generate $jj$-coupling shell-model wave functions 
from $\alpha$ cluster models.
Here, $jj$-coupling shell model wave function
of $\nuc{Li}{9}$ is regarded as a cluster, which corresponds to the subclosure 
configuration of $p_{3/2}$ for the neutrons, and we discuss how
the two neutrons connect two $\nuc{Li}{9}$ clusters.
Until now, 
most of the clusters in the conventional models have been limited to the closures
of the three-dimensional harmonic oscillators, such as $\nuc{He}{4}$, $^{16}$O, and $^{40}$Ca;
however, owing to AQCM, it is feasible to utilize the $jj$-coupling shell model wave functions as 
plural subsystems quite easily.
The appearance of a rotational band structure with a cluster structure around 
the four-body threshold energy is discussed.
\end{abstract}
\maketitle
%
%%%%%%%%%%%%%%%%% 
\section{Introduction}
%%%%%%%%%%%%%%%%% 
\par
\par
The $\nuc{He}{4}$ nuclei have been known to have large binding energy 
(28.3~MeV)
distinctly in the light mass region.
In contrast, it is known that the relative interaction between $\nuc{He}{4}$ nuclei is weak.
Therefore, they can serve as subsystems 
in some of the light nuclei,
called $\alpha$ cluster structure~\cite{Brink,RevModPhys.90.035004}.
The $\alpha$ cluster structure has been studied for decades,
and one of the most famous examples is the second $0^+$ state of $\nuc{C}{12}$
with  a developed three-$\alpha$ cluster structure~\cite{Hoyle1954,FREER20141},
which is called the Hoyle state.
Many cluster models have proven to be capable of describing various properties of the Hoyle state~\cite{PTPS.68.29,PhysRevLett.87.192501}.
\par
In most of the conventional cluster models, however, 
only the states corresponding to the closure
of the three-dimensional harmonic oscillator, such as $\nuc{He}{4}$, $\nuc{O}{16}$, and $\nuc{Ca}{40}$,
have been treated as subsystems called clusters.
In these cases, unfortunately,
the contribution of the non-central interactions (spin-orbit and tensor interactions),
which are quite important in the nuclear systems, 
vanishes after the antisymmetrization of the wave functions.
This is the consequence of the fact 
that the closure configurations of the major shells are spin-zero systems,
and the non-central interactions do not contribute to such spinless systems. 
In actual nuclear systems, on the contrary, the contribution of the spin-orbit  interaction
is essential, which enables to
explain the observed magic numbers;
the subclosure configurations of 
the $jj$-coupling shell model
($f_{7/2}$, $g_{9/2}$, and $h_{11/2}$)
correspond to the observed magic numbers of $28$, $50$, and $126$~\cite{Mayer}.
If we enlarge the model space of the cluster models
and open the path to another symmetry,
indeed this spin-orbit interaction 
works as a driving force to break the
$\alpha$ clusters, for instance in $\nuc{C}{12}$~\cite{PhysRevC.70.054307}.
\par
Therefore, the important task is to include the spin-orbit contribution
by extending the traditional cluster models;
we proposed the antisymmetrized quasi cluster model
(AQCM)~\cite{PhysRevC.94.064324,PhysRevC.73.034310,PhysRevC.75.054309,PhysRevC.79.034308,PhysRevC.83.014302,PhysRevC.87.054334,ptep093D01,ptep063D01,ptepptx161,PhysRevC.97.014307,PhysRevC.98.044306,PhysRevC.101.034304,PhysRevC.102.024332}.
This method allows the smooth transformation of the $\alpha$ cluster model wave functions to
$jj$-coupling shell model ones.
We call the clusters that feel the spin-orbit contribution 
after this transformation
quasi clusters.
The conventional $\alpha$ cluster models cover the model space of closure of major shells
(corresponding to the magic numbers of 2, 8, and 20),
and now
the subclosure configurations of the $jj$-coupling shell model,
$p_{3/2}$, $d_{5/2}$, $f_{7/2}$, $g_{9/2}$ $etc.$,
are covered by our AQCM~\cite{ptep093D01}.
\par
The achievement of AQCM allows us to use  $jj$-coupling shell model wave functions
as subsystems of the nuclei, which is the beginning of a new cluster model.
At first, we have shown the possibility of 
$\nuc{C}{14}$, $\nuc{He}{6}$, and $\nuc{Li}{9}$
as clusters~\cite{PhysRevC.101.034304,PhysRevC.102.024332},
where subclosure configuration of $p_{3/2}$ plays an important role.
The cluster structures
of 
$\nuc{Be}{16}$ ($\nuc{He}{8}$+$\nuc{He}{8}$),
$\nuc{B}{17}$  ($\nuc{He}{8}$+$\nuc{Li}{9}$),
$\nuc{C}{18}$   ($\nuc{Li}{9}$+$\nuc{Li}{9}$),
$\nuc{C}{24}$ (three $\nuc{He}{8}$),
and
$\nuc{Ar}{42}$ (three $\nuc{C}{14}$),
have been investigated around the corresponding threshold energies.
\par
In this study, we further add neutrons and discuss  
in $\nuc{C}{20}$ the $\nuc{Li}{9}$+$\nuc{Li}{9}$ cluster configuration
as an example.
It has been known 
in $\nuc{Li}{11}$
that the two neutrons have neutron halo structure around $\nuc{Li}{9}$~\cite{PhysRevLett.55.2676}.
It is intriguing to see how such structure is affected when 
another $\nuc{Li}{9}$ approaches to the halo neutrons.
In the Be isotopes, it has been extensively discussed that neutrons
perform molecular-orbital motion around $\alpha$ clusters;
very developed $\alpha$-$\alpha$ cluster structure appears
when two neutrons occupy the $\sigma$ orbit~\cite{PhysRevC.61.044306,PhysRevLett.100.182502}.
This idea can be extended to the linear-chain structure of three $\alpha$ clusters
in the C isotopes~\cite{PhysRevC.64.014301,ZIM.15}.
If the neutrons in the halo state perform molecular orbital motion around two $\nuc{Li}{9}$ clusters
with large distances, it paves the way to a novel binding mechanism in the excited states of the neutron-rich nuclei.
\par
This paper is organized as follows. 
 The framework is described in  Sec.~\ref{Frame}.
The results are shown in Sec.~\ref{Results}.
 The conclusions are presented in Sec.~\ref{Concl}.
%
%%%%%%%%%%%%%%%%%%%%%%%%%%%%%%%%%%% 
\section{framework}
\label{Frame}
%%%%%%%%%%%%%%%%%%%%%%%%%%%%%%%%%%% 
%
\subsection{Basic feature of AQCM}
\par
AQCM 
allows the smooth transformation of the cluster model 
wave functions to the $jj$-coupling shell model ones.
In AQCM, each single particle is described by a Gaussian form
as in many other cluster models including the Brink model~\cite{Brink},
\begin{equation}	
  \phi^{\tau, \sigma} \left( \bm{r} \right)
  =
  \left(  \frac{2\nu}{\pi} \right)^{\frac{3}{4}} 
  \exp \left[- \nu \left(\bm{r} - \bm{\zeta} \right)^{2} \right] \chi^{\tau,\sigma}, 
  \label{spwf} 
\end{equation}
where the Gaussian center parameter $\bm{\zeta}$
is related to the expectation 
value of the position of the nucleon,
and $\chi^{\tau,\sigma}$ is the spin-isospin part of the wave function.
For the size parameter $\nu$, 
here we use $\nu = 0.20 \, \mathrm{fm}^{-2}$, which gives the optimal $0^+$ energy
of $\nuc{C}{12}$ within a single AQCM basis state.
The Slater determinant is constructed from 
these single-particle wave functions by antisymmetrizing them.
\par
Next, we focus on the Gaussian center parameters
$\left\{ \bm{\zeta}_i \right\}$.
As in other cluster models, here four single-particle 
wave functions with different spin and isospin
sharing a common 
$\bm{\zeta}$ value correspond to an $\alpha$ cluster.
This cluster wave function is transformed into
$jj$-coupling shell model based on the AQCM.
When the original value of the Gaussian center parameter $\bm{\zeta}$
is $\bm{R}$,
which is 
real and
related to the spatial position of this nucleon, 
it is transformed 
by adding the imaginary part as
\begin{equation}
  \bm{\zeta} = \bm{R} + i \Lambda \bm{e}^{\text{spin}} \times \bm{R}, 
  \label{AQCM}
\end{equation}
where $\bm{e}^{\text{spin}}$ is a unit vector for the intrinsic-spin orientation of this
nucleon. 
The control parameter $\Lambda$ is associated with the breaking of the cluster,
and with a finite value of $\Lambda$, the two nucleons with opposite spin orientations 
have the $\bm{\zeta}$ values, which are complex conjugate  with each other.
This situation corresponds to the time-reversal motion of two nucleons.
After this transformation, the $\alpha$ clusters are called quasi clusters.
We can generally create 
the single-particle orbits of the $jj$-coupling shell model
by taking the limits of $\bm{R} \to 0$ and $\Lambda \to 1$.

\subsection{Wave function for $\nuc{C}{20}$}
\par
The total wave function for $\nuc{C}{20}$ is the superposition of different Slater determinants, $\{ \Phi_i \}$, 
\begin{equation}
  \Psi_{J^\pi} = \sum_{i,K} c^K_i P_{J^\pi}^K \Phi_i.
\end{equation}
All Slater determinants are projected to the eigen states of parity and angular momentum by
using the projection operator $P_{J^\pi}^K$,
\begin{equation}
  P_{J^\pi}^K
  =
  P^\pi \frac{2J+1}{8\pi^2}
  \int d\Omega \, {D_{MK}^J}^* R \left(\Omega \right).
\end{equation}
Here ${D_{MK}^J}$ is the Wigner $D$-function 
and $R\left(\Omega\right)$ is the rotation operator
for the spatial and spin parts of the wave function.
This integration over the Euler angle $\Omega$ is numerically performed.
The operator $P^\pi$ is for the parity projection ($P^\pi = \left(1+P^r\right) / \sqrt{2}$ for
the positive-parity states, where $P^r$ is the parity-inversion operator), 
which is also performed numerically.
The coefficients $\left\{ c^K_i \right\}$ are obtained together with the energy eigenvalue $E$
when we diagonalize the norm and Hamiltonian ($H$) matrices, namely  
by solving the Hill-Wheeler equation.
\begin{eqnarray}
 \sum_{i,j,K,K'} && (<\Phi_i | (P_{J^\pi}^{K'})^\dagger H P_{J^\pi}^K | \Phi_j>  \nonumber \\
&&- E <\Phi_i | (P_{J^\pi}^{K'})^\dagger P_{J^\pi}^K | \Phi_j>)c^K_j 
 = 0.
\end{eqnarray}
This angular momentum projection enables to generate different $K$ number states
as independent basis states
from each Slater determinant.
\par
Each Slater determinant consists of the antisymmetrized product of single-particle wave functions. 
\begin{eqnarray}	
\Phi_i = {\cal A} \{ &&\phi^{\tau_1, \sigma_1} \left( \bm{r}_1,\bm{\zeta}_1 \right)
\phi^{\tau_2, \sigma_2} \left( \bm{r}_2,\bm{\zeta}_2 \right)
 \cdots \nonumber \\
&& \cdots 
\phi^{\tau_{19}, \sigma_{19}} \left( \bm{r}_{19},\bm{\zeta}_{19} \right)
\phi^{\tau_{20}, \sigma_{20}} \left( \bm{r}_{20},\bm{\zeta}_{20} \right)
 \}_i.
\end{eqnarray}
Here, the single-particle wave functions
from $\phi^{\tau_1, \sigma_1}\left( \bm{r}_1,\bm{\zeta}_1 \right)$ 
   to $\phi^{\tau_9, \sigma_9}\left( \bm{r}_9,\bm{\zeta}_9 \right)$
belong to one $\nuc{Li}{9}$ cluster,
whereas those from $\phi^{\tau_{10}, \sigma_{10}}\left( \bm{r}_{10},\bm{\zeta}_{10} \right)$ 
to $\phi^{\tau_{18}, \sigma_{18}}\left( \bm{r}_{18},\bm{\zeta}_{18} \right)$ are for another
$\nuc{Li}{9}$. 
For each $\nuc{Li}{9}$ cluster, we introduce the subclosure configuration
of $(s_{1/2})^{2}(p_{3/2})^4$ for the neutrons.
For the proton part, 
the last protons in two $\nuc{Li}{9}$ clusters are introduced as time-reversal configurations,
$ \left| 3/2\ 3/2 \right\rangle $ and $ \left| 3/2\ -3/2 \right\rangle $.
Then these two $\nuc{Li}{9}$ clusters are separated with the relative distance of $d_i$,
which is randomly generated with equal probability between 0.5~fm and 5.0~fm.

\par
The two valence neutrons, 
$\phi^{\tau_{19}, \sigma_{19}}\left( \bm{r}_{19},\bm{\zeta}_{19} \right)$
and  $\phi^{\tau_{20}, \sigma_{20}}\left( \bm{r}_{20},\bm{\zeta}_{20} \right)$, 
are introduced to have opposite spin directions (spin-up and spin-down).
Their Gaussian center parameters ($\bm{\zeta}_{19}$ and $\bm{\zeta}_{20}$) 
are generated by using random numbers $\left\{ r_k \right\}$,
which have the probability distribution $P\left(\left|r_k\right|\right)$
proportional to  $\exp \left[-\left|r_k\right| / \sigma \right]$,
\begin{equation}
  P \left( \left| r_k \right| \right) \propto \exp \left[- \left| r_k \right| / \eta \right].
\end{equation}
The value of $\eta$ is chosen to be $1.5\,\mathrm{fm} $
which corresponds to the standard deviation of 3.67~fm.
After generating $\left\{ r_k \right\}$, we multiply the sign factor to each $r_k$,
which allows $r_k$ to be positive and negative with equal probability.
The resultant random numbers are applied to all three ($x$, $y$, $z$) components 
of the Gaussian center parameters of the two valence neutrons.

\subsection{Hamiltonian}
\par
The Hamiltonian consists of the kinetic energy and 
potential energy terms.
For the potential part, the interaction consists of the central 
($\hat{V}_{\text{central}}$), 
spin-orbit
($\hat{V}_{\text{spin-orbit}}$), 
and 
Coulomb terms. For the central part, the Tohsaki interaction~\cite{PhysRevC.49.1814}  
is adopted. This interaction has finite ranges for the three-body terms in 
addition to two-body terms, which is designed to reproduce both saturation 
properties and scattering phase shifts of two $\alpha$ clusters. For the spin-orbit part, 
we use the spin-orbit term of the G3RS interaction~\cite{PTP.39.91}, which is a realistic 
interaction originally  developed to reproduce the nucleon-nucleon scattering phase 
shifts.
\par
The Tohsaki interaction 
consists of two-body ($V^{\text{(2)}}$)  and three-body ($V^{\text{(3)}}$) terms:
\begin{equation}
  \hat{V}_{\text{central}}
  =
  \frac{1}{2} \sum_{i \neq j} V^{\text{(2)}}_{ij} 
  +
  \frac{1}{6} \sum_{i \neq j, j \neq k, i \neq k}  V^{\text{(3)}}_{ijk},
\end{equation}
where $V^{\text{(2)}}_{ij}$ and $V^{\text{(3)}}_{ijk}$ have three ranges,
\begin{align}
  V^{\text{(2)}}_{ij}
  = & \,
      \sum_{\alpha=1}^3
      V^{\text{(2)}}_\alpha
      \exp\left[- \frac{\left(\vec{r}_i - \vec{r}_j \right)^2}{\mu_\alpha^2} \right]
      \left(W^{\text{(2)}}_\alpha - M^{\text{(2)}}_\alpha P^\sigma P^\tau \right)_{ij},
      \label{2body} \\
  V^{\text{(3)}}_{ijk}
  = & \,
      \sum_{\alpha=1}^3
      V^{\text{(3)}}_\alpha 
      \exp\left[- \frac{\left(\vec r_i - \vec r_j \right)^2}{\mu_\alpha^2}
      -                                                     
      \frac{\left(\vec r_i - \vec r_k\right)^2}{\mu_\alpha^2} \right]
      \notag \\
  & \times 
    \left( W_\alpha^{\text{(3)}} - M_\alpha^{\text{(3)}} P^\sigma P^\tau \right)_{ij} 
    \left( W_\alpha^{\text{(3)}} - M_\alpha^{\text{(3)}} P^\sigma P^\tau \right)_{ik}.
\end{align}
Here, $P^\sigma P^\tau$ represents the exchange of the spin-isospin part of the wave functions
of interacting two nucleons.
The physical coordinate for the $i$th nucleon is $\vec{r}_i$.
The details of the parameters are shown in Ref.~\cite{PhysRevC.49.1814},
but we use F1' parameter set for the Majorana parameter ($M_\alpha^{\text{(3)}}$) of the 
three-body part introduced in Ref.~\cite{PhysRevC.94.064324}.
\par
The G3RS interaction~\cite{PTP.39.91}
is a realistic interaction, and the spin-orbit term has the following form;
\begin{equation}
  \hat{V}_{\text{spin-orbit}}
  =
  \frac{1}{2} \sum_{i \ne j} V^{ls}_{ij}, 
\end{equation}
where
\begin{equation}
  V^{ls}_{ij}
  =
  \left(
    V_{ls}^1 e^{-d_{1} \left( \vec{r}_i - \vec{r}_j \right)^{2}}
    -
    V_{ls}^2 e^{-d_{2} \left( \vec{r}_i - \vec{r}_j \right)^{2}} \right) 
  P\left({}^{3}O\right)
  {\vec{L}}\cdot{\vec{S}}.
  \label{Vls}
\end{equation}
Here, $\vec{L}$ is the angular momentum for the relative motion
between the $i$th and $j$th nucleons, and $\vec{S}$ is the sum of the spin operator
for these two interacting nucleons.
The operator $P\left({}^{3}O\right)$ stands for the projection onto the triplet-odd state.
The strength of the spin-orbit interactions is set to $V_{ls}^1=V_{ls}^2=1800 \, \mathrm{MeV}$,
which allows consistent description of $\nuc{C}{12}$ and $\nuc{O}{16}$~\cite{PhysRevC.94.064324}.
\section{Results}
\label{Results}
\subsection{Energy convergence of $\nuc{C}{20}$}
%
%%%%%%%%%%%%%%%%%%%%%%%%%%%%%%%%%%%%%%%%%%%%%%%% 
\begin{figure}
  \centering
  \includegraphics[width=5.5cm]{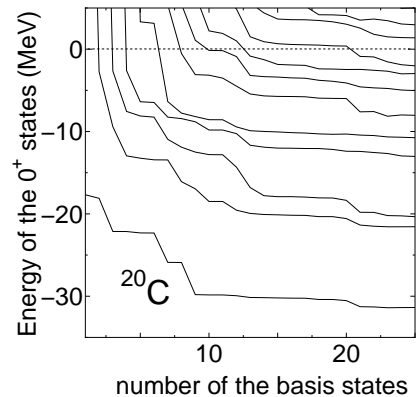} 
  \caption{
    Energy curves for the  $0^+$ states
    of $\nuc{C}{20}$ calculated with the $\nuc{Li}{9}$+$\nuc{Li}{9}$+$n$+$n$ model.
    Horizontal axis
    shows the number of Slater determinants superposed. 
    Energies are measured from the four-body threshold.}
  \label{c20-ec}
\end{figure}
%%%%%%%%%%%%%%%%%%%%%%%%%%%%%%%%%%%%%%%%%%%%%%%% 
\par
We start the discussion
with the $0^+$ energy curves
of $\nuc{C}{20}$ calculated with the present $\nuc{Li}{9}$+$\nuc{Li}{9}$+$n$+$n$ model.
The horizontal axis 
of Fig.~\ref{c20-ec} 
shows the number of Slater determinants superposed, and 
for each Slater determinant,
the distances between two $\nuc{Li}{9}$ 
($d$)
and the positions for the Gaussian center parameters
for the two valence neutrons 
($\bm{\zeta}_{19}$ and $\bm{\zeta}_{20}$)
are randomly generated.
The energies are measured from the four-body threshold.
This calculation is based on the bound state approximation and it could be possible that
some of the obtained states above the neutron threshold
are continuum states, but the states with flat energies
after rapid convergence are candidates for the resonance states.
The lowest $0^+$ state converges to $-31.4$~MeV, 
and experimentally, the ground state of $\nuc{C}{20}$ is located at $-28.5$~MeV from the threshold.
The obtained $0^+$ energies are summarized in Table~\ref{properties}.
\subsection{Principal quantum number of the harmonic oscillator}
%%%%%%%%%%%%%%%%%%%%%%%%%%%%%%%%%%%%%%%%%%%%%%%% 
\begin{table}[ht]
\caption{
Properties of 
$0^+$ states of $\nuc{C}{20}$
obtained by superposing Slater determinants.
The energies are
measured from the $\nuc{Li}{9}+\nuc{Li}{9}+n+n$ threshold ($E$ (MeV)).
Expectation values of the principal quantum number 
for the proton part ($N$-proton)  and neutron part ($N$-neutron) are listed
together with those of the single-particle parity of the protons ($spp$-proton) and neutrons ($spp$-neutron).}
\label{c20-n-table}
\centering
\begin{tabular}{rrrrrr}
\hline 
\hline
 & $E$ (MeV) & $N$-proton & $N$-neutron & $spp$-proton & $spp$-neutron\\
\hline
 1 & $-31.37$ & $4.19$ & $18.61$ & $-1.90$  & $1.85$  \\
 2 & $-21.56$ & $4.43$ & $18.92$ & $-1.78$  & $1.87$  \\
 3 & $-20.34$ & $4.90$ & $19.82$ & $-1.56$  & $1.64$  \\
 4 & $-13.02$ & $4.83$ & $19.86$ & $-1.58$  & $1.06$  \\
 5 & $-10.75$ & $5.60$ & $21.41$ & $-1.27$  & $-0.79$  \\
 6 & $-8.03$  & $4.71$ & $19.62$ & $-1.59$  & $1.49$  \\
 7 & $-5.03$  & $5.68$ & $21.44$ & $-1.29$  & $1.04$  \\
 8 & $-2.96$  & $6.12$ & $21.79$ & $-0.35$  & $0.08$  \\
 9 & $-2.02$  & $5.77$ & $21.68$ & $-1.13$  & $1.15$  \\
10 & $1.38$   & $5.96$ & $22.13$ & $-1.09$  & $0.89$  \\
\hline
\hline
\label{properties}
\end{tabular}
\end{table}
%%%%%%%%%%%%%%%%%%%%%%%%%%%%%%%%%%%%%%%%%%%%%%%%%
\par
We discuss the property of each state obtained after superposing the basis states.
Particular focus is placed on the point
which states have the character of $\nuc{Li}{9}+\nuc{Li}{9}$ cluster configuration. 
One of the physical quantities, which characterize each state,
is the expectation value of the principal quantum number $\hat{N}$ of the harmonic oscillator, 
\begin{equation}
  \hat{N} = \sum_i \bm{a}^\dagger_i \cdot \bm{a}_i.
\end{equation}
The summation can be taken independently for the proton part and neutron part.
In Table~\ref{properties}, the column ``$N$-proton'' stands for
the expectation value of $\hat{N}$
for the protons and ``$N$-neutron'' for the neutrons.
The lowest Pauli-allowed values are 4 for the protons (two are in the $s$-shell and four are in the $p$-shell)
and 18 for the neutrons (two are in the $s$-shell, six are in the $p$-shell, and six are in the $sd$-shell).
The lowest $0^+$ state has the values of 4.19 (protons) and 18.61 (neutrons) fairly close to the lowest Pauli-allowed values,
which indicates that the state has pure shell-model character.
However, the excited states have much larger values.
\par
\begin{figure}
  \centering
  \includegraphics[width=5.5cm]{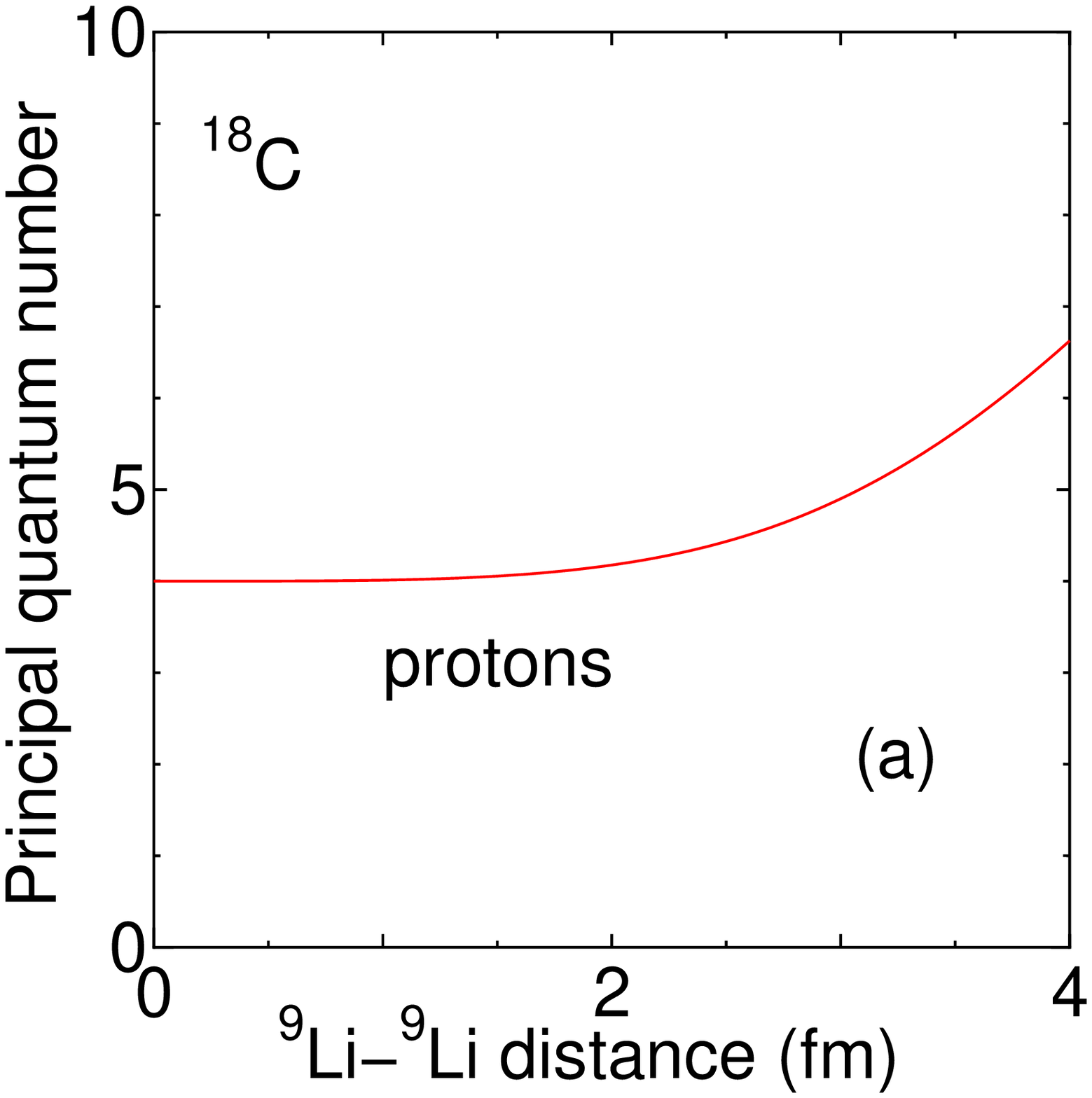} 
   \includegraphics[width=5.5cm]{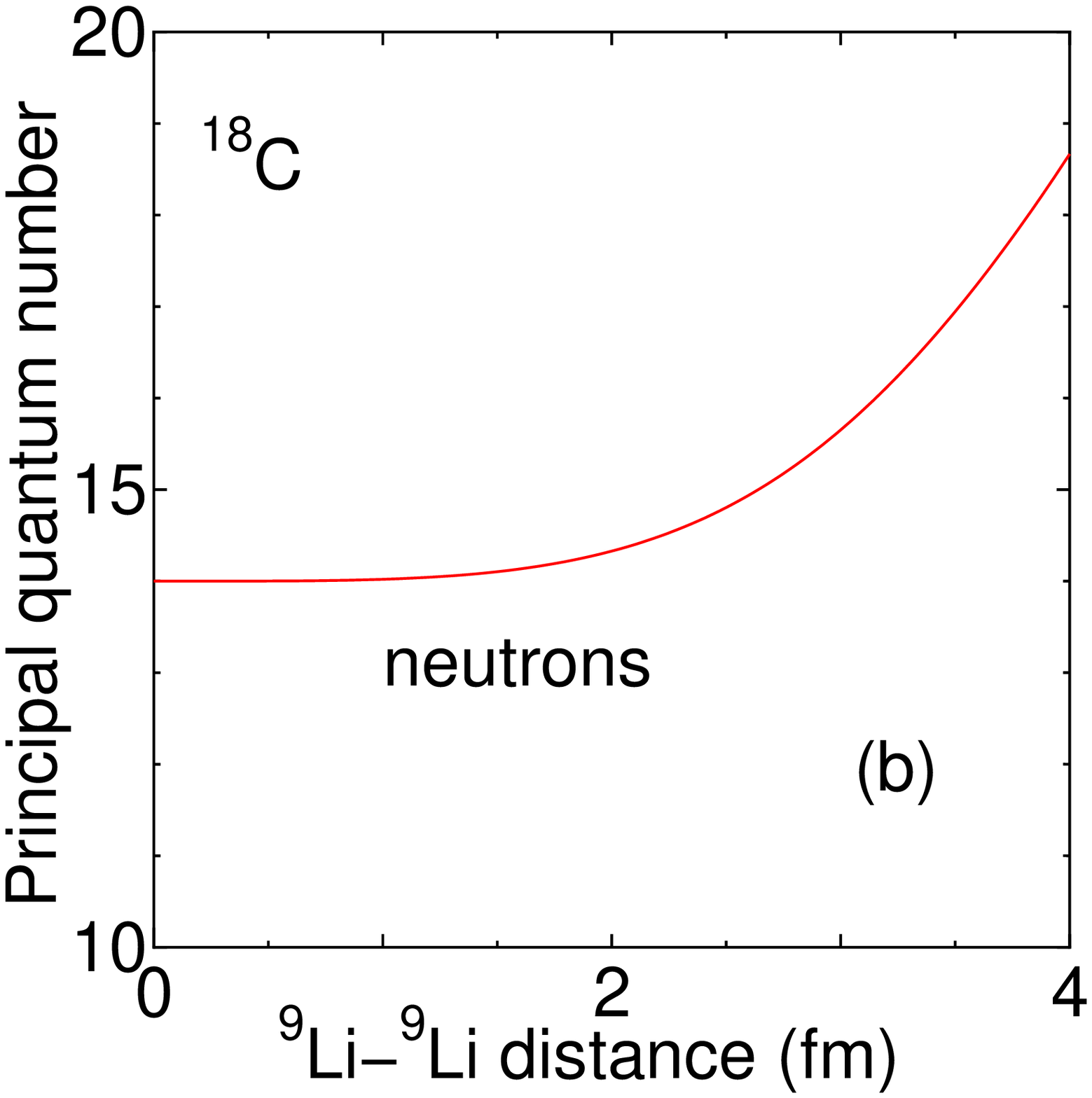} 
  \caption{
    Principal quantum number $N$ of the harmonic oscillator for the $0^+$ state of $\nuc{C}{18}$ as a function of the distance between two $\nuc{Li}{9}$ clusters.
    (a): protons and (b): neutrons.}
  \label{c18-n}
\end{figure}

It should be stressed that the principal quantum number $N$ for the protons 
($N$-proton) is governed by
the distance between two $\nuc{Li}{9}$ clusters,
since protons are only in the $\nuc{Li}{9}$ clusters and the internal wave functions
of each $\nuc{Li}{9}$ is frozen except for the antisymmetrization effect.
Thus, 
the $N$-proton values listed in Table~\ref{properties}
is considered to contain information for 
the distance between two $\nuc{Li}{9}$ clusters,
although the expectation values are obtained after superposing
Slater determinants with different $\nuc{Li}{9}$-$\nuc{Li}{9}$ distances.
To extract
the contribution of the  $\nuc{Li}{9}+\nuc{Li}{9}$ core part,
in Fig.~\ref{c18-n}, we show the $N$ values of $\nuc{C}{18}$ 
without the two valence neutrons as a function of the distance between two $\nuc{Li}{9}$ clusters
(Fig.~\ref{c18-n}(a): protons and Fig.~\ref{c18-n}(b): neutrons). 
At the zero-limit 
for the distance $d$ between two $\nuc{Li}{9}$,
the values are 4 (protons) and 14 (neutrons),
The values increase to 4.9 (protons) and 15.7 (neutrons) at 
$d = 3$~fm, and they grow to 5.6 and 16.9 at $d = 3.5$~fm 
(they further increase to 6.6 and 18 at $d = 4$~fm).
\par
According to Table~\ref{properties},
the fifth state at $-10.75$~MeV, seventh state at $-5.03$~MeV, eighth state at $-2.96$~MeV,
ninth state at $-2.02$~MeV, and tenth state at $1.38$~MeV
have the $N$-proton values of $5.5-6.0$.
Thus, it is quite likely that these states have 
the cluster structure with the $\nuc{Li}{9}$-$\nuc{Li}{9}$ distances of $3.5-4.0$~fm.
Among them, the eighth state at $-2.96$~MeV
has the largest $N$-proton value of 6.12,
and we consider this state as a candidate for the cluster state
representing these states.
The $N$-proton value of 6.12 corresponds to the $\nuc{Li}{9}$-$\nuc{Li}{9}$ distances of $3.75$~fm
in $\nuc{C}{18}$ without the valence neutrons (Fig.~\ref{c18-n}(a)).
In our preceding work for $\nuc{C}{18}$~\cite{PhysRevC.102.024332},
$\nuc{Li}{9}+\nuc{Li}{9}$ cluster state has been shown to
appear slightly below the corresponding threshold energy. 
Therefore, it is rather reasonable to find here the appearance of
the cluster state in $\nuc{C}{20}$ after adding two valence neutrons
to $\nuc{C}{18}$ below the four-body
threshold.
This state shows rapid energy convergence in Fig~\ref{c20-ec}, 
which is considered to be a possible resonance state.
\par
If we assume that the $0^+$ state of $\nuc{C}{20}$ at $-2.96$~MeV,
which is the candidates for the cluster state,  
has the $\nuc{Li}{9}$-$\nuc{Li}{9}$ distances of about 3.75~fm,
the neutrons in the $\nuc{Li}{9}+\nuc{Li}{9}$ core part 
must have the $N$-neutron 
value around 17.75 (Fig.~\ref{c18-n}(b)).
Therefore, we can deduce the contribution of the two valence neutrons
in the state at $-2.96$~MeV by subtracting  17.75
from the $N$-neutron value of 21.79
listed in Table~\ref{properties}.
The two valence neutrons
are estimated to have the $N$ value of about 4.0 in the cluster state, 
and each valence neutron shares the value of $\sim$2.0.
Thus, it turns out that valence neutrons remain in two-node orbits
as in the ground state.
%%%%%%%%%%%%%%%%%%%%%%%%%%%%%%%%%%%%%%%%%%%%%%%% 
%%%%%%%%%%%%%%%%%%%%%%%%%%%%%%%%%%%%%%%%%%%%%%%% 
\subsection{Single-particle parity}
\par
To discuss the properties of the two valence neutrons in $\nuc{C}{20}$ in more detail,
next we calculate the single-particle parity.  In this framework, the single-particle orbits introduced 
are non-orthogonal, and therefore
we cannot directly discuss the parity of each orbit. 
Another difficulty of directly discussing the parity of each single-particle
comes from the superposition of Slater determinants performed here. 
Nevertheless, we can get
insights for the single-particle parity from the expectation value of 
the single-particle operator,
which is the sum of the parity inversion operator for each nucleon~\cite{PhysRevC.74.067304},
\begin{equation}
  \hat{O}^{spp}=\sum_{i} P^r_i.
\end{equation}
Here $P^r_i$ is the parity-inversion operator for the $i$-th nucleon.
The eigenvalues of $P^r_i$ is 1 and $-1$, for the positive-parity orbit and negative-parity orbit, respectively.
The summation can be taken over the protons or neutrons independently.
\par
We calculate the expectation value of this single-particle parity for the $0^+$ states of $\nuc{C}{20}$
listed in Table~\ref{properties},
which are 
obtained by superposing Slates determinants.
In Table~\ref{properties}, the column ``$spp$-proton'' is for
the protons and ``$spp$-neutron'' is for the neutrons.
The ground state has the values of $-1.90$ (protons) and 1.85 (neutrons)
indicating that the state has the character of the lowest shell-model state.
In the lowest shell-model state,
two protons and two neutrons
are in the $s$-shell with positive-parity, 
four protons and six neutrons are in the $p$-shell with negative-parity,
and six neutrons are in the $sd$-shell with positive-parity, 
and hence, the single-particle parity becomes $2-4=-2$ 
for the protons and $2-6+6=2$ for the neutrons.
However, the candidates for the cluster state
(eighth states at $-2.96$~MeV in Table~\ref{properties}
identified as cluster state in the previous subsection)
has quite different values; 
$-1.09$ ($spp$-proton) and 0.08 ($spp$-neutron). 
\par
\begin{figure}
  \centering
  \includegraphics[width=5.5cm]{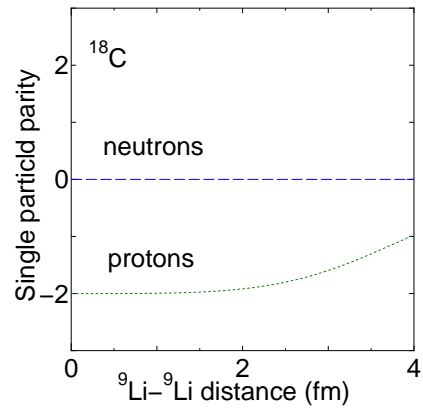} 
  \caption{
Single-particle parity for the $0^+$ state of $\nuc{C}{18}$ as a function of the distance between two $\nuc{Li}{9}$ clusters.
Dotted line is for the protons and dashed-line is for the neutrons.
}
  \label{c18-spp}
\end{figure}

In the present model, the protons are only in the $\nuc{Li}{9}+\nuc{Li}{9}$ core part
as mentioned before.
Thus, the calculated single-particle parity of the protons
($spp$-proton)
is governed by
the distance
between two $\nuc{Li}{9}$ clusters.
The correspondence between the distance between two $\nuc{Li}{9}$ clusters
and $spp$-proton can be clarified in $\nuc{C}{18}$ by removing the two valence neutrons.
Figure~\ref{c18-spp}
depicts the single-particle parity for the $0^+$ state of $\nuc{C}{18}$ as a function of the distance between two $\nuc{Li}{9}$ clusters, where
the dotted line is for the protons ($spp$-proton) and the dashed-line is for the neutrons ($spp$-neutron).
For the protons (dotted line), the value is around $-2$ at small relative distances
because the last proton in each $\nuc{Li}{9}$ occupies $p$-shell-like one-node orbit.
However the value 
starts slightly deviating from $-2$ with increasing the distance,
suggesting partial excitation of the protons to two-node orbits with positive-parity, 
while the total parity of the two-$\nuc{Li}{9}$ system is always projected to positive.
The value for the protons is $-1.60$ at the $d = 3.5$~fm
and it becomes $-0.97$ at $d = 4$~fm. 
\par
We can compare this result with the $spp$-proton values of $\nuc{C}{20}$ listed in Table~\ref{properties},
where the two valence neutrons are added and the states with different $\nuc{Li}{9}$-$\nuc{Li}{9}$
distances are superposed.
In the previous subsection, we have identified that the eighth state
in Table~\ref{properties} as the candidates for the cluster state with the 
$\nuc{Li}{9}$-$\nuc{Li}{9}$ distances of $\sim$$3.75$~fm.
The obtained $spp$-proton values of $\nuc{C}{20}$
is $-0.35$ for the eighth state, 
quite different from $-2$.
The obtained value 
is consistent with 
that for the states having finite $\nuc{Li}{9}$-$\nuc{Li}{9}$ distances,
suggesting the prominent cluster structure.
\par
Next, we discuss the single-particle parity of the neutron part.
It is intriguing to point out that in $\nuc{C}{18}$ without the two valence neutrons, 
the single-particle parity of the neutrons  
is always zero independent of the $\nuc{Li}{9}$-$\nuc{Li}{9}$ distance (Fig.~\ref{c18-spp} dashed line).
This is coming from the fact that the neutrons in both $\nuc{Li}{9}$ 
have identical (subclosure) configurations except for the central positions
of the clusters. The antisymmetrization effect allows
the linear combination of the single-particle orbits, and the combinations of
the two orbits with good parity around the center of left or right cluster 
 always create a pair of positive-parity and negative-parity orbits.
Thus, the single-particle parity for the neutrons of $\nuc{C}{18}$ is zero,
and in $\nuc{C}{20}$, the $spp$-neutron values purely show
the contribution of the two valence neutrons. 
In Table~\ref{properties}, the ground state and low excited states have the value close to 2
indicating that the two valence neutrons are in the $sd$-shell-like positive-parity orbits.
However, the candidate
for the cluster state 
(eighth state in Table~\ref{properties})
have the value of 0.08
completely different from 2. 
The result indicates that the valence neutrons have more complex characters than 
``pure two-node orbits'', which was not evident when we discussed the principal quantum number $N$ in the previous subsection.
One of the possible explanations is that the orbits of the valence neutrons 
have mixed components of the $p$-shell, $sd$-shell, and $pf$-shell orbits.

%%%%%%%%%%%%%%%%%%%%%%%%%%%%%%%%%%%%%%%%%%%%%%%%% 
%%%%%%%%%%%%%%%%%%%%%%%%%%%%%%%%%%%%%%%%%%%%%%%%

\subsection{Rotational band structure of $\nuc{C}{20}$} 
\begin{figure}
  \centering
  \includegraphics[width=5.5cm]{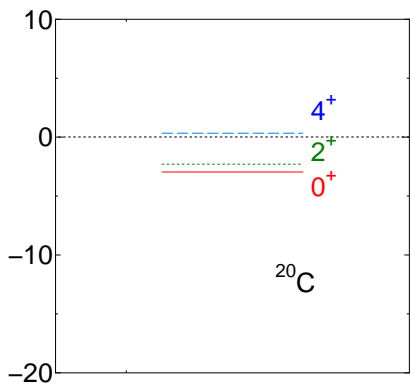} 
  \caption{
Rotational band structure of $\nuc{C}{20}$
measured from the $\nuc{Li}{9}+\nuc{Li}{9}+n+n$ threshold energy.
Solid, dotted, and dashed lines are for $0^+$, $2^+$, and $4^+$ states.
}
  \label{c20-rb}
\end{figure}

The same calculation can be performed also for the $2^+$ and $4^+$ states
of $\nuc{C}{20}$.
The eighth states in Table~\ref{properties} at $-2.96$~MeV,
which is a candidate for the cluster structure, serves as a band head state of 
a rotational band structure
as shown in Fig.~\ref{c20-rb},
where the solid, dotted, and dashed lines are for $0^+$, $2^+$, and $4^+$ states, respectively.
The $2^+$ state at $-2.29$~MeV and $4^+$ states at $0.33$~MeV 
are connected with the $0^+$ state at $-2.96$~MeV and
classified into a rotational band judging from the 
similarity of the calculated properties (principal quantum number and single-particle parity)
and electromagnetic transition 
probabilities among the states.
We have not considered the energy of each state in this classification,
but eventually, typical level spacings of the rotational band structure comes out
around the $\nuc{Li}{9}+\nuc{Li}{9}+n+n$ threshold energy.
The rotational band shows small level spacings between $0^+$ and $2^+$
reflecting the large deformation. 
%%%%%%%%%%%%%%%%%%%%%%%%%%%%%%%%%%%%%%%%%%%%%%%% 
%%%%%%%%%%%%%%%%%%%%%%%%%%%%%%%%%%%%%%%%%%%%%%%%%

%%%%%%%%%%%%%%%%% 
\section{Conclusions} 
\label{Concl}
%%%%%%%%%%%%%%%%% 
\par
In this study, we discussed the structure of $\nuc{C}{20}$ by 
introducing a four-body cluster model with $\nuc{Li}{9}+\nuc{Li}{9}+n+n$ configurations.
The recent development of the antisymmetrized quasi cluster model (AQCM) 
enables us to utilize $jj$-coupling shell-model wave functions 
as plural subsystems quite easily.
Until now, many works have shown in $\nuc{Li}{11}$ the presence of the halo structure
comprised of the weakly bound two neutrons around $\nuc{Li}{9}$, 
and our study was motivated by a question 
how this halo structure changes when another $\nuc{Li}{9}$ approaches.
\par
In our preceding work for $\nuc{C}{18}$~\cite{PhysRevC.102.024332}, it has been discussed that
$\nuc{Li}{9}+\nuc{Li}{9}$ cluster state appears below the threshold, and therefore, it is natural in $\nuc{C}{20}$ 
to see the appearance of
the cluster states below the threshold after adding the two valence neutrons.
The candidate for the cluster state was identified by calculating 
the principal quantum number and single-particle parity.
The candidate for the cluster state serves as a band head state and
forms a rotational band structure, where
the members are collected by the 
similarity of the properties 
and electromagnetic transition 
probabilities among the states.
The states have small level spacings 
reflecting the large deformation.
\par
As future work, we investigate the 
possibility of molecular-orbital states.
In the present analysis, we discussed that atomic-orbit states,
where neutron(s) sticks to one of the two centers, but
in the molecular-orbit state, 
each valence neutron rotates around two centers equally with good parity.
Such molecular-obit states are expected to appear around the threshold energy.
Also, we will connect the calculation to the nuclear reaction and discuss
how we can populate these states in the actual experiment.

\begin{acknowledgments}
  The authors would like to thank the discussion with Dr.~M.~Sasano (RIKEN).
  The numerical calculations have been performed using the computer facility of 
  Yukawa Institute for Theoretical Physics,
  Kyoto University. 
\end{acknowledgments}
\bibliography{biblio_ni.bib}
% \bibliography{references22.bib}
%
\end{document}